\documentstyle[psfig,twocolumn,prl,aps]{revtex}

\begin{document}

\draft

\title{Pseudogap and Conduction Dimensionalities in High-$T_c$ Superconductors}

\author{Andrew Das Arulsamy$^{a,b,}$\thanks{E-mail: sadwerdna@hotmail.com}}

\address{$^a$Department of Physics, \
Center for Superconducting and Magnetic Materials, \
Faculty of Science, \
National University of Singapore, \
2 Science Drive 3, \
117542 Singapore, Singapore \\} 

\address{$^b$No. 22, Jalan Melur 14, \
Taman Melur, 68000 Ampang, \
Selangor DE, \
Malaysia \\} 

\maketitle

%\tightenlines

\begin{abstract}
The nature of normal state charge-carriers' dynamics and the transition in conduction and gap dimensionalities between 2D and 3D for YBa$_2$Cu$_3$O$_{7-\delta}$ and Bi$_2$Sr$_2$Ca$_{1-x}$Y$_x$Cu$_2$O$_8$ high-$T_c$ superconductors were described by computing and fitting the resistivity curves, $\rho(T,\delta,x)$.  These were carried out by utilizing the 2D and 3D Fermi liquid (FL) and ionization energy ($E_I$) based resistivity models coupled with charge-spin (CS) separation based {\it t-J} model [Phys. Rev. B {\bf 64}, 104516 (2001)]. $\rho(T,\delta,x)$ curves of Y123 and Bi2212 samples indicate the beginning of the transition of conduction and gap from 2D to 3D with reduction in oxygen content (7$-\delta$) and Ca$^{2+}$ (1$-x$) as such, $c$-axis pseudogap could be a different phenomenon from superconductor and spin gaps. These models also indicate that the recent MgB$_2$ superconductor is at least not Y123 or Bi2212 type.  
\end{abstract}


\section{Introduction}

Two-dimensional (2D) charge-carriers' transport properties in normal state of high-$T_c$ superconductors (HTSC) in accordance with FL theory were widely employed in order to describe the normal state resistivities such as $\rho_{ab}(T)$, $\rho_c(T)$ and related phenomena\cite{moriya1}.  Experimental evidences supporting the existence of Fermi surfaces in superconducting cuprates have also been obtained with various techniques\cite{peter2}.  One of the well established and intriguing aspect is the coexistence of metallic and semiconducting resistivities in both $\rho_{ab}(T)$ and $\rho_c(T)$. In addition to such anomaly, transition of 2D normal state transport properties in HTSC to a 3D one equivalent to semiconductors with doping is an equally interesting aspect to analyze.  In the attempt to derive a $c$-axis resistivity model in accordance with 2D FL theory\cite{arulsamy3}, electrons were believed to tunnel through Cu-O$_2$ planes as required by the resonating-valence-bond (RVB) theory\cite{anderson4} in which it has to overcome an anomalous gap\cite{zheng5} which is currently termed as the Pseudogap ($\Delta_{PG}$)\cite{timusk6}. The models derived in ref.\cite{arulsamy3} emphasized on the variation and the importance of $E_I$ as a $c$-axis anomalous gap in HTSC. Lately, it has been shown that the opening of spin gap ($\Delta_{SG}$) reduces spin scattering hence, $\rho_{ab}(T)$ reduces rapidly than $T$-linear below $T^*$ (characteristics temperature), as a consequence $\rho_c(T)$ is supposed to reveal semiconducting behavior below $T^*$\cite{anderson4,timusk6}. Interestingly, Takenaka {\it et al}.\cite{takenaka7} also claimed that superconducting fluctuations are unlikely in HTSC since the fluctuations should be proportional to $T$.  However, one should note that semiconducting behavior of $\rho_c(T)$ below $T_{crossover}$ for underdoped samples or at least for BSCCO\cite{takenaka7} does not coincide with opening of $\Delta_{SG}$ at $T^*$ in $\rho_{ab}(T)$, which put further constraints on $\Delta_{SG}$ and $\Delta_{PG}$.  Recently, Krasnov {\it et al}.\cite{krasnov8} have found inconsistencies between superconductor gap ($\Delta_{SCG}$) and $\Delta_{PG}$ in term of $H$ and $T$ dependence of each gap where $\Delta_{PG}$ persists below $T_c$ and is independent of $H$ and $T$ whereas $\Delta_{SCG}$ closes at $H$ $\to$ $H_{c2}(T)$ and $T$ $\to$ $T_c$. These phenomena point towards a different origin for both gaps.  Parallel to this, Renner {\it et al}.\cite{renner9} concluded that reduction of $\Delta_{PG}$ $\to$ 0 at $T^*$ in various experiments is to be considered as a characteristic energy scale and not as $T$ where $\Delta_{PG}$ $\to$ 0. They further proved the existence of $\Delta_{PG}$ for all under-, over- and optimally-doped samples.  On the other hand, $\rho(T)$ transition occurring from metallic (normal state) superconductor (MS) to metallic-semiconducting crossover (normal state; with and without $T^*$) superconductor (MSS) and finally to non-superconducting semiconductor (NS) or vice versa with doping is another critical sub-aspect of the above-mentioned conduction peculiarities.  The above stated incompatibilities are highlighted and described herein by first deriving a 3D phenomenological resistivity model that fulfills FL theory to accommodate NS characteristics. I.e., this is an extension to 2D $c$-axis resistivity model that accommodates normal state characteristics of MSS without $T^*$ and semiconducting (normal state) superconductor (SS)\cite{arulsamy3}. Apart from that, $\Delta_{SG}$ is assumed to occur only in $ab$-planes at $\Delta_{PG}$ $<$ $\Delta_{SG}$ while $\Delta_{PG}$ is the $c$-axis gap.  These assumptions are valid since weak dependence of $\Delta_{SG}$ with $H$ and $T$\cite{krasnov8}, rapid reduction of $\rho_{ab}(T)$ with decreasing $T$ below $T^*$\cite{takenaka7,ito10}, incompatibility between $T^*$ ($ab$-planes) and $T_{crossover}$ ($c$-axis)\cite{takenaka7} and the confinement behavior\cite{anderson4} of $\rho(T)$ in $ab$-planes and $c$-axis may account for a separate origin of $\Delta_{SG}$ and $\Delta_{PG}$.  Thus, 2D $\Delta_{PG}$ in $c$-axis normal state conduction that could be thought of as a function of doping, be it ionic substitution or oxygen content, extends to $ab$-planes and becomes 3D due to increasing $\Delta_{PG}$ with decreasing oxygen content or certain ionic substitution. Such properties in Y123\cite{leridon11} and Bi2212\cite{ruan12} samples with different oxygen content (7$-\delta$) and Ca$^{2+}$ (1$-x$) are studied. Another possibility to further enhance the mentioned transition with persistent proof is that by introducing other researcher's experimental results in a similar context. Hence the results obtained by Watanabe {\it et al.}\cite{takenaka7}, Ito {\it et al.}\cite{ito10} and Hagen {\it et al.}\cite{hagen13} are utilized and coupled with above-mentioned results to illustrate and justify the normal state conduction as well as $\Delta_{PG}$'s dimensionality transition with oxygen content and Y$^{3+}$. 

Ichinose {\it et al}.\cite{ichinose14} using effective gauge-field theory of {\it t-J} model (GFT) in CS separated (CSS) state have derived an expression for $\rho_{ab}(T)$ at $T$ $<$ $T^*$, which is given by

\begin{eqnarray}
\rho_{ab}(T) & = & BT[1-C(T^* - T)^d]. \label{eq:1} 
\end{eqnarray}

$B$ and $C$ are constants, dependent on holons' scattering rate ($\tau_h$) and doping respectively while $d(\delta,T)$ being the critical exponent in 2D. Actually, $d$ characterizes the gauge field mass, $m_A$ in such a way that $m_A$ $\propto$ $(T^* - T)^d$. Moreover, $d$-wave spinon pairing parameter, $\lambda$ also $\propto$ $(T^* - T)^d$ for $T$ $<$ $T^*$ hence, $\rho_{ab}(T)$ = $BT$ at $T$ $>$ $T^*$ since $\lambda$ = 0\cite{ichinose14}. In summary, this work differs from previous studies and new in the sense that $E_I$ based $\rho(T)$ model is introduced to account for $\Delta_{PG}$ and $T_{crossover}$ between metallic (at high $T$) to semiconducting (at low $T$) behaviors as well as evaluates quantitatively the transition of $\rho(T)$ and $\Delta_{PG}$ dimensionalities. These were carried out by employing 2D and 3D FL and RVB-GFT models. Moreover, the possibility of separate origins for $\Delta_{PG}$ and $\Delta_{SG}$ is proposed in order to resolve  $\Delta_{PG}$ and $\Delta_{SCG}$ inconsistencies that states $\Delta_{PG}$ and $\Delta_{SCG}$ could share a similar origin\cite{kleefisch15}.

\section{Resistivity models in the Normal State of HTSC}

The 3D $E_I$ based $\rho(T)$ model can be derived as follows; electron's distribution as a function of average $E_I$ as an anomalous constraint has been derived from Fermi-Dirac (FD) statistics and is given by\cite{arulsamy3}

\begin{eqnarray}
f_e(E) &=& exp\left[\frac{E_F - E_I - E}{k_BT}\right]. \label{eq:2}
\end{eqnarray}

Similarly, the probability function for holes is 

\begin{eqnarray}
f_h(E) &=& exp\left[\frac{E - E_I - E_F}{k_BT}\right]. \label{eq:3}
\end{eqnarray}

Both functions (Eq.~\ref{eq:2} and Eq.~\ref{eq:3}) were derived by inserting conditions, $E_{electron}$ = $E_{initial state}$ + $E_I$ and $E_{hole}$ = $E_{initial state}$ $-$ $E_I$ respectively. This is to justify that an electron to occupy a higher state $N$ from initial state $M$ is more probable than from initial state $L$ if condition $E_I(M)$ $<$ $E_I(L)$ at certain temperature, $T$ is satisfied. As for a hole to occupy a lower state $M$ from initial state $N$ is more probable than to occupy state $L$ if the same condition above is satisfied. $E_{initial state}$ is the energy of a particle in a given system at a certain initial state and ranges from $+\infty$ to 0 for electrons and 0 to $-\infty$ for holes. Existence of energy gap, $E_g$ that tied to lattice is unclear thus it is (if any) coupled with $\Delta_{PG}$ while $E_I$ is tied to ions. Note that above distributions can be reduced to FD distributions if $E_I$ $\rightarrow$ 0. Subsequently, the respective normal state cocentration of electrons and holes can be obtained from

\begin{eqnarray}
n &=& \int^{\infty}_0{f_e(E)N_e(E)dE},
\label{eq:4}
\end{eqnarray}

\begin{eqnarray}
p &=& \int_{-\infty}^0{f_h(E)N_h(E)dE}.
\label{eq:5}
\end{eqnarray}

$N_e(E)$ and $N_h(E)$ are electron's and hole's density of states respectively. Substituting Eq.~\ref{eq:2}, Eq.~\ref{eq:3} and $N_{e,h}$($E$,3D) = $[2E(m_{e,h}^*)^3]^{1/2}/\pi^2\hbar^3$ accordingly into Eq.~\ref{eq:4} and Eq.~\ref{eq:5}, and solving those integrals give the respective 3D concentration of electrons and holes as 

\begin{eqnarray}
n &=& 2\left[\frac{m^*_ek_BT}{2\pi\hbar^2}\right]^\frac{3}{2}exp\left[\frac{E_F - E_I}{k_BT}\right], \label{eq:6}
\end{eqnarray}

\begin{eqnarray}
p &=& 2\left[\frac{m^*_hk_BT}{2\pi\hbar^2}\right]^\frac{3}{2}exp\left[\frac{- E_F - E_I}{k_BT}\right]. \label{eq:7}
\end{eqnarray}

where $k_B$ is Boltzmann constant, $m^*_e$ and $m^*_h$ are effective electrons' and holes' masses respectively, $\hbar$ = $h/2\pi$, $h$ is Plank constant and $E_F$ is Fermi level. In addition, Eq.~\ref{eq:6} and Eq.~\ref{eq:7} can be written in the form of geometric mean, which is given by 

\begin{eqnarray}
\sqrt{np} &=& 2\left[\frac{k_BT}{2\pi\hbar^2}\right]^\frac{3}{2}(m^*_em^*_h)^\frac{3}{4}exp\left[\frac{- E_I}{k_BT}\right]. \label{eq:8}
\end{eqnarray}

Incorporating, $m$ = $m^*_e$ = $m^*_h$, $n$ = $\sqrt{np}$ (Eq.~\ref{eq:8}) and $1/\tau$ = $A_2T^2$ into the metallic resistivity equation which is defined as $\rho$ = $m/ne^2\tau$ with $e$ and $\tau$ being electron's charge and scattering rate respectively, then one can derive the 3D phenomenological $\rho(T,E_I)$ model for NS samples as given below

\begin{eqnarray}
\rho(T,E_I) & = & \frac{A_2}{2e^2}\left[\frac{2\pi\hbar^2}{k_B}\right]^{\frac{-3}{2}}(m^*_e m^*_h)^{\frac{-3}{4}}\sqrt{T} exp\left[\frac{E_I}{k_BT}\right], \label{eq:9}
\end{eqnarray}

$A_2$ being a constant independent of $T$ and $\tau$ dependent. The 2D $c$-axis $\rho(T)$ model derived elsewhere is given by\cite{arulsamy3}

\begin{eqnarray}
\rho_c(T,E_I) & = & A\frac{\pi\hbar^2}{e^2k_B}T exp\left[\frac{E_I - E_F}{k_BT}\right]. \label{eq:10}
\end{eqnarray}

Note that $A$ also $\tau$ dependent, $\Delta_{PG}$ = $E_I - E_F$ = $k_BT_{crossover}$ for $c$-axis in 2D HTSC and {\bf $\Delta_{PG}$} = $E_I$ for 3D NS. The 2D $\Delta_{PG}$ for $c$-axis conduction is solely dependent on electron-ion attraction and it is different from semiconductor gap, which is due to energy band splitting or lattice based gap. In CSS state, $\rho_{ab}(T)$ and $\rho_c(T)$ are interrelated\cite{anderson4}, thus Eq.~\ref{eq:1} and Eq.~\ref{eq:10} can be summed accordingly to accomodate the process spinon($s$) + holon($h$) $\rightleftharpoons$ electron($e$) to represent the coupling between $c$-axis and $ab$-plane conduction. Electrons in $c$-axis indeed has different physical properties than spinons and holons in $ab$-planes, therefore two different models (FL and GFT) are needed to describe the so-called normal state charge carriers' properties of HTSC. As such, these two models have to be coupled in order to obtain the overall behaviour due to $s$ + $h$ $\rightleftharpoons$ $e$ and 2D confinement. The Ioffe-Larkin\cite{ichinose14} formula in $ab$-planes is given by $\sigma^{-1}_{total}$ = $\sigma^{-1}_h$ + $\sigma^{-1}_s$ where $\sigma^{-1}_{ab}$ = $\sigma^{-1}_h$ i.e., the total resistivity in $ab$-planes is equals to holons only since spinons conductivity diverges $\sigma_s$ $\to$ $\infty$ as a result of $\Delta_{SG}$ condensation\cite{ichinose14} below $T^*$. The total resistivity in $c$-axis and $ab$-planes namely, $\rho_c^{total}(T)$ ($\rho_c^{(t)}(T)$) and $\rho_{ab}^{(t)}(T)$ in CSS region can be respectively written as, 

\begin{eqnarray}
\rho_c^{(t)}(T) &=& \sigma^{-1}_e + \sigma^{-1}_{e \rightleftharpoons s + h} \label{eq:11} 
\end{eqnarray}

\begin{eqnarray}
\rho_{ab}^{(t)}(T) &=& \sigma^{-1}_s + \sigma^{-1}_h + \sigma^{-1}_{s + h \to e} \label{eq:12} 
\end{eqnarray}

The term $\sigma^{-1}_{e \rightleftharpoons s + h}$ in Eq.~\ref{eq:11} is defined to be the resistivity caused by blockage in the process $e$ $\rightleftharpoons$ $s$ + $h$ or the resistivity caused by the blockage faced by electrons to enter the $ab$-planes ($e$ $\to$ $s$ + $h$) and the blockage faced by spinons and holons to leave the $ab$-planes ($s$ + $h$ $\to$ $e$). These blockages originate from the non-spontaneity conversion of $e$ $\rightleftharpoons$ $s$ + $h$. The resistivity, $\sigma^{-1}_{e \rightleftharpoons s + h}$ can also be solely due to the blockage of $e$ $\to$ $s$ + $h$ or $s$ + $h$ $\to$ $e$. Actually, if the magnitude of blockage in $e$ $\to$ $s$ + $h$ $>$ $s$ + $h$ $\to$ $e$ then the blockage of $e$ $\to$ $s$ + $h$ contributes to $\rho_c^{(t)}(T)$ apart from $\sigma^{-1}_e(T)$ or vice versa if $e$ $\to$ $s$ + $h$ $<$ $s$ + $h$ $\to$ $e$. In short, if one of the conversion, say $e$ $\to$ $s$ + $h$ is less spontaneous than $s$ + $h$ $\to$ $e$, then the former conversion determines the $\sigma^{-1}_{e \rightleftharpoons s + h}$. Moreover, increment in $\sigma^{-1}_{ab}$ further blocks $e$ $\rightleftharpoons$ $s$ + $h$ that leads to a larger $\sigma^{-1}_{e \rightleftharpoons s + h}$ hence, $\sigma^{-1}_{e \rightleftharpoons s + h}$ $\propto$ $\sigma^{-1}_{ab}$. This proportionality can be interpreted as the additional scattering for the electrons to pass across $ab$-planes. If $e$ $\rightleftharpoons$ $s$ + $h$ is spontaneous then $\sigma^{-1}_{e \rightleftharpoons s + h}$ or the magnitude of the corresponding blockage equals 0. The term $\sigma^{-1}_{s + h \to e}$ in Eq.~\ref{eq:12} is defined to be the resistivity caused by the process $s$ + $h$ $\to$ $e$ occuring in the $ab$-planes. Alternatively, it is the resistivity caused by the electrons in $ab$-planes. If $s$ + $h$ $\to$ $e$ is completely blocked in $ab$-planes then $\sigma^{-1}_{s + h \to e}$ = 0. Any increment in $\sigma^{-1}_e$ also increases $\sigma^{-1}_{s + h \to e}$ therefore $\sigma^{-1}_{s + h \to e}$ $\propto$ $\sigma^{-1}_e$. This latter interesting effect will be discussed further in Sec. III with experimental proofs in term of $\Delta_{PG}$, $T^*$ and doping. Consequently, by employing the above arguments, Eq.~\ref{eq:11} and Eq.~\ref{eq:12} can be written as 

\begin{eqnarray}
\rho_c^{(t)}(T) &=& \sigma^{-1}_e + \beta[\sigma^{-1}_h + \sigma^{-1}_s] = \rho_c + \beta\rho_{ab} \label{eq:13} 
\end{eqnarray}

\begin{eqnarray}
\rho_{ab}^{(t)}(T) &=& \sigma^{-1}_s + \sigma^{-1}_h + \gamma\sigma^{-1}_e =  \rho_{ab} + \gamma\rho_c \label{eq:14} 
\end{eqnarray}

$\beta$ and $\gamma$ are constants of proportionality, which are tightly related to the degree of contribution from $ab$-planes and $c$-axis respectively. Assume that, if $s$ and $h$ in $ab$-planes of HTSC are independent of $e$ in $c$-axis ($s$ + $h$ $\rightleftharpoons$ $e$ is invalid), then there will be no contribution of $c$-axis conduction into $ab$-planes or vice versa as a result of 2D confinement, in which $\beta$ = $\gamma$ = 0. However this is not so in actual HTSC, where $\rho_e$ is dominant in $c$-axis with minor contribution from $ab$-planes and $\rho_{s+h}$ is dominant in $ab$-planes with minor contribution from $c$-axis due to 2D confinement and also as a result of $s$ + $h$ $\rightleftharpoons$ $e$. If $s$ + $h$ $\rightleftharpoons$ $e$ exists in 3D system, meaning $s$, $h$ and $e$ can conduct in any directions then $\rho$ = $\rho_s$ + $\rho_h$ + $\rho_e$ where $\beta$ = $\gamma$ = 1. Parallel to this, one has $\rho(T)$ at $T$ $<$ $T^*$ for HTSC in the form of
 
\begin{eqnarray}
\rho(T) & = & ATexp\left[\frac{\Delta_{PG}}{T}\right] + BT[1 - C(T^* - T)^d], \label{eq:15}
\end{eqnarray}

subsequently, at $T$ $>$ $T^*$, $\rho(T)$ can be written as 

\begin{eqnarray}
\rho(T) & = & ATexp\left[\frac{\Delta_{PG}}{T}\right] + BT. \label{eq:16}
\end{eqnarray}

Note that $A$ as in Eq.~\ref{eq:15} and Eq.~\ref{eq:16} are also a function of $\gamma$ with $\gamma$ $<$ 1 whereas $\beta$ incorporated in $B$ equals 1 if $\rho(T)$ = $\rho_{ab}^{(t)}(T)$. If $\rho(T)$ = $\rho_{c}^{(t)}(T)$, then $\beta$ is less than 1 and $\gamma$ = 1. $\rho^{(t)}(T)$ in $c$-axis or $ab$-planes or in pure thin films is only valid in CSS region. In non-CSS region, Eq.~\ref{eq:10} is solely for $c$-axis and $\rho_{ab}(T)$ = $BT$, assuming the system is still 2D. In other words, the process $s$ + $h$ $\rightleftharpoons$ $e$ is invalid in non-CSS region and gives $\beta$ = $\gamma$ = 0 if the system is still 2D (charge carriers in $ab$-planes are independent of charge carriers in $c$-axis). In 3D, regardless of $s$ + $h$ $\rightleftharpoons$ $e$, $\rho(T)$ = $\rho_c(T)$ = $\rho_{ab}(T)$ that suggests resistivities in all directions are identical as long as 2D confinement is invalid. In both $ab$-planes and $c$-axis, $T$-linear metallic behavior will recover at $T$ $>$ $\Delta_{PG}$. It is stressed that the analysis of the coupled Eq.~\ref{eq:13} and Eq.~\ref{eq:14} in term of $\delta$, $T_{crossover}$ and $T^*$ are to further enhance the viability of these models to capture various $\rho^{thin films}(T)$, $\rho_c^{(t)}(T)$ and $\rho_{ab}^{(t)}(T)$ curves in the normal state reported thus far apart from studying the transition in $\Delta_{PG}$ and conduction dimensionalities.

\section{Discussion}

It is clear that slightly underdoped Y123 sample (A1)\cite{leridon11} as shown in Fig.~\ref{fig:1} exposes $T$-linear metallic normal state at $T$ $>$ $T^*$ with $T^*$ = 196.5 K ($\Delta_{PG}$ $<$ $T_c$ $<$ $T^*$) whereas sample B2 (YBa$_2$Cu$_3$O$_{7-\delta}$, $\delta$ = 0.39) from Carrington {\it et al.}\cite{leridon11} has $T^*$ (275.8 K) above $T_c$ and also shows a prior indication of $T_{crossover}$ since $\rho(T)$'s slope decreases near $T_c$ where $T_{crossover}$ is probably below $T_c$ hence, the upturn is not observed. Sample C3 of Bi$_2$Sr$_2$Ca$_{1-x}$Y$_{x}$Cu$_2$O$_8$ ($x$ = 0.05) is metallic with $T^*$ = 201.0 K whereas D4 ($x$ = 0.11) has both linear-metallic above $T^*$ (246.4 K) and semiconducting below $T_{crossover}$ (72.0 K) behaviors as depicted in Fig.~\ref{fig:2} ($T_c$ $<$ $\Delta_{PG}$ $<$ $T^*$). I.e., sample D4 is equivalent to MSS with $T^*$. The plots in Fig.~\ref{fig:1} for A1 and B2 and Fig.~\ref{fig:2} for C3 and D4 are fitted using Eq.~\ref{eq:9}, Eq.~\ref{eq:15} and  Eq.~\ref{eq:16} in order to accurately distinguish and identify the discrepancies between 2D and 3D conductions. However, fitting of Eq.~\ref{eq:9} ($\Box$) is only shown for sample C3 for the sake of clarity. Note also that $A_2$ from Eq.~\ref{eq:9} equals $(A_2/2e^2)(2\pi\hbar^2/k_B)^{-3/2}(m^*_em^*_h)^{-3/4}$ and $A$ from Eq.~\ref{eq:15} and Eq.~\ref{eq:16} equals $A\gamma\pi\hbar^2/e^2k_B$ for the sake of convenience. All $T^*$ were determined from calculated fittings due to its uncertainty from $\rho(T,\delta,x)$ curves\cite{leridon11}. The fitted parameters are as follows: sample A1; $\Delta_{PG}$(2D) = 30 K, $A$ = 1.4, $B$ = 9.9 $\times$ $10^{-8}$, $C$ = 10, $d$ = 2.75, $\Delta_{PG}$(3D) = $-$85.2 K, $A_2$ = 35.62, sample B2; $\Delta_{PG}$(2D) = 40 K, $A$ = 1.4065, $B$ = 2.5 $\times$ $10^{-8}$, $C$ = 100/3, $d$ = 2.68, $\Delta_{PG}$(3D) = $-$110 K, $A_2$ = 40.2, sample C3; $\Delta_{PG}$(2D) = 55 K, $A$ = 2.545 $\times$ $10^{-3}$, $B$ = 1.9 $\times$ $10^{-11}$, $C$ = 25, $d$ = 3.0, $\Delta_{PG}$(3D) = $-$57 K, $A_2$ = 6.39 $\times$ $10^{-2}$, sample D4; $\Delta_{PG}$ (2D) = 65 K, $A$ = 4.2 $\times$ $10^{-3}$, $B$ = 9.9 $\times$ $10^{-11}$, $C$ = 125, $d$ = 2.4, $\Delta_{PG}$(3D) = $-$213 K, $A_2$ = 20.2 $\times$ $10^{-2}$. Fittings in $T$ $>$ $T^*$ region give $AT$ $\gg$ $BT$ hence, the latter term of Eq.~\ref{eq:16} is neglected. The readers are referred to Ref.\cite{arulsamy3} for a detailed discussion on the ionic substitution effect on $\rho(T)$ and $T_{crossover}$. Calculated values of $\Delta_{PG}$ (3D) is $<$ 0 for all samples (A1, B2, C3 and D4) where it does not hold any physical meaning and could indicate a non-3D conduction. Basically, these negative values are necessary for fittings of Eq.~\ref{eq:9} (3D) to compensate small $d\rho/dT$ due to $\rho(T)$ $\propto$ $\sqrt{T}$, which is less than $d\rho/dT$ ($\rho(T)$ $\propto$ $T$) from Eq.~\ref{eq:16} (2D). As such, Eq.~\ref{eq:9} (3D) is unsuitable for $T$-linear curves' characterizations as in HTSC. In addition, fittings with Eq.~\ref{eq:9} are not exactly $T$-linear metallic above $T^*$ that also can be easily verified from Eq.~\ref{eq:9} ($\rho(T)$ $\propto$ $\sqrt{T}$; this non-$T$-linear proportionality from Eq.~\ref{eq:9} itself is the main criterion that invalidates 3D conduction). It is also important to realize that $\Delta_{PG}$ obtained via resistivity measurements or charge carriers' dynamics always gives a relatively low value since conduction's path are as such where the carriers path are along low $\Delta_{PG}$. On the contrary ARPES or any other tunneling techniques probe the 2D area throughout the materials without any discrimination of high or low $\Delta_{PG}$ and in average, it is always higher than one obtained from resistivity measurements. It seems that reduction in oxygen content and Y$^{3+}$ substitution increases $\Delta_{PG}$. As anticipated, $\rho_c(T)$ for crystalline Y123 HTSC as given in Ref.\cite{hagen13} cannot be fitted with Eq.~\ref{eq:9} (3D) even with reasonable accuracy whereas it can be easily fitted with Eq.~\ref{eq:10} (2D)\cite{arulsamy3} that could somewhat proves 2D conduction in the normal state of HTSC. Separately, resistivity curve of MgB$_2$ as given in Ref.\cite{pradhan16} cannot be fitted with both Eq.~\ref{eq:16} and Eq.~\ref{eq:9} that could suggest the normal state of MgB$_2$ is neither in CSS state nor in a gapped fermion state respectively. This is somehow expected since there are no $T^{1/2}$ and metallic $T$-linear behaviors, $T^*$ or $T_{crossover}$ and the measured gap also closes at $T$ $=$ $T_c$ without any indication of a $\Delta_{PG}$ in the normal state\cite{karapetrov17}. Furthermore, $\rho(T)$'s curve for MgB$_2$ has been fitted with Bloch-Gruneisen equation~\cite{pradhan16} with rather good fitting by Tu {\it et al.} that suggests the normal state of this material is identical to conventional superconductors. It is found that Eq.~\ref{eq:15} and Eq.~\ref{eq:16} account very well for changes occuring in $\rho(T)$ with $\Delta_{PG}$, oxygen content, $T^*$, $T_{crossover}$ and ionic substitutions ($x$). Apart from that, it also points out that $\Delta_{PG}$ has an independent origin from $\Delta_{SCG}$ in accordance with recent experimental evidences\cite{krasnov8,renner9}. However, there are also contradicting evidences\cite{kleefisch15} that claim $\Delta_{PG}$ evolves from $\Delta_{SCG}$ in which, this issue will be tackled in the following paragraph. 

Two of the unique properties of the normal state $\rho_{ab}(T)$ and $\rho_c(T)$ are gapless behavior of charge-carriers in $ab$-planes and a $\Delta_{PG}$ induced conduction in $c$-axis. Recall that the former generalization was partly interpreted from ref.\cite{anderson4}. The two completely conflicting-to-each-other conduction has enabled one to distinguish and break down those two resistivities in two separate axes that may lead one to suggest that $c$-axis $\Delta_{PG}$ is indeed 2D. It is believed that certain ionic substitution or reduction in oxygen content in MS sample would increase $\Delta_{PG}$ and as a consequence, extends to $ab$-planes i.e., the conduction and gap become 3D (NS). According to RVB theory, this dimensionality transition is possible if a large number of spinons and holons ceased to become real electrons in $c$-axis since $\Delta_{PG}$ is too large for tunneling. Therefore, accumulation of spinons and holons could now lead to electrons' formation in $ab$-planes ($s$ + $h$ $\to$ $e$), that would eventually leads to the transition of $\Delta_{PG}$'s dimensionality from 2D $\to$ 3D since $\Delta_{PG}$ develops with the formation of electrons in $ab$-planes. This electrons' formation in $ab$-planes at low $T$ below $\Delta_{PG}$ could be interpreted as superconducting fluctuations. It has been shown experimentally that reduction in oxygen content or certain ionic substitution leads 2D HTSC $\to$ 3D non-superconducting semiconductors\cite{fisher18}. Initial accumulation of spinons and holons would give rise to $\Delta_{SG}$ and $T^*$ prior to electrons' formation in $ab$-planes. Interestingly, one should carefully note that initial reduction of oxygen content as reported in Refs.\cite{takenaka7,ito10} for BSCCO$_{8+\delta}$ and YBCO$_{7-\delta}$ single crystals respectively increases the magnitude of $\Delta_{PG}$ therefore, less electrons are able to conduct in $c$-axis and as a consequence more spinons and holons are being formed in $ab$-planes. This signifies that $\Delta_{SG}$ increases initially with increasing of $s$ and $h$ concentrations ($n_s$ and $n_h$), hence this explains the increment of $T^*$ accordingly. Further reduction in oxygen content however, would lead to semiconductor behavior at a certain $T_{crossover}$ since $\Delta_{PG}$ is too large for electrons to conduct in $c$-axis. This effect give rise to over-accumulation of spinons and holons and subsequently leads to the formation of electrons ($s$ + $h$ $\to$ $e$) in $ab$-planes that could be interpreted as superconducting fluctuations as mentioned previously. Simultaneously $\Delta_{PG}$ (originally existed in $c$-axis) opens up in $ab$-planes due to $s$ + $h$ $\to$ $e$ in $ab$-planes as pointed out by the 3$^{rd}$ term in Eq.~\ref{eq:12}. This can be clearly seen in sample D4 below $T_{crossover}$ (72.0 K) due to Y$^{3+}$ substitution where $\Delta_{PG}$ have increased from 55 K for $x$ = 0.05 to 65 K for $x$ = 0.11. Specifically, one can evaluate the average $E_I$ of Y$^{3+}$ and Ca$^{2+}$ as 1260 KJmol$^{-1}$ $>$ 867 KJmol$^{-1}$ respectively\cite{arulsamy3} for Bi$_2$Sr$_2$Ca$_{1-x}$Y$_x$Cu$_2$O$_8$ (sample C3 and D4). In short, $\Delta_{PG}$ and conduction dimensionalities increase towards a 3D with decreasing oxygen content and increasing Y$^{3+}$ or with certain ionic substitution that has higher ionization energy relatively. Add to that, development of $\Delta_{PG}$ in $ab$-planes also accounts for the disappearance of $T^*$ since electrons formation reduces $n_s$ and $n_h$. Simply put, $\Delta_{SG}$ that varies with spinon pairing is independent of $\Delta_{PG}$ that varies with $E_I$. This may indicate the $\Delta_{SG}$ that closes at $T^*$ could also evolve from $\Delta_{SCG}$ when the temperature is raised above $T_c$ ($\ll$ $T^*$). In contrast, $\Delta_{PG}$ as a different entity could exist at all temperature range.    

\section{Conclusions}

Consistent with above arguments, one may conclude that 2D normal state conduction with charge-carriers' confinement in separate axes namely, $ab$-planes and $c$-axis is vital as far as HTSC are concerned. I.e., this confinement leads to 2D conduction (MS and MSS) and gaps ($\Delta_{PG}$ and $\Delta_{SG}$) where the transition to 3D (NS) with certain ionic doping or reduction in oxygen content suppresses the superconducting characteristics. As such, $\Delta_{PG}$ and $\Delta_{SG}$ could be different entities where $\Delta_{SG}$ that tied to spinon-pairing may evolve or scales with $\Delta_{SCG}$. Besides, the respective total resistivities in $c$-axis and $ab$-planes are given by $\rho_c^{(t)}(T)$ = $\rho_c(T,E_I)$ + $\beta\rho_{ab}(T)$ and $\rho_{ab}^{(t)}(T)$ = $\rho_{ab}(T)$ + $\gamma\rho_c(T,E_I)$. $\beta$ and $\gamma$ represents partial contribution from $ab$-planes and $c$-axis respectively or due to coupling effect ($e$ $\rightleftharpoons$ $s$ + $h$) in CSS state. Resistivities in these forms are able to characterize both $T^*$ and $T_{crossover}$ in $ab$-plane as well as $T_{crossover}$ in $c$-axis. It has been shown both quantitatively and qualitatively that the effect of oxygen doping and ionic substitution on the total resistivities are in accordance with FL-RVB-GFT framework. It is also found that doping increases $\Delta_{PG}$ in such a way that it extends to $ab$-planes and leads to the transition of $\Delta_{PG}$ and conduction dimensionalities from 2D $\to$ 3D. On the other hand, these models (Eq.~\ref{eq:9}, Eq.~\ref{eq:15} and Eq.~\ref{eq:16}) via fittings suggest that MgB$_2$'s normal state characteristics could not be identical to cuprate based HTSC or at least not similar to Y123 or Bi2212\cite{bud'ko19}.  

\section*{ACKNOWLEDGMENTS}

ADA would like to thank the National University of Singapore and Physics department for the financial assistance. The author is also grateful to A. Innasimuthu, I. Sebastiammal, A. Das Anthony and Cecily Arokiam for their partial financial assistances and also to P. C. Ong and M. T. Ong for sharing their unpublished experimental data and matlab program respectively.

\begin{figure}

\caption {Experimental $\rho(T,\delta)$ curves for A1 ($\diamond$) and B2 ($\triangle$) samples that have been fitted using Eq.~\ref{eq:15} and Eq.~\ref{eq:16} (solid lines) at $T$ $<$ $T^*$ and $T$ $>$ $T^*$ respectively. Both A1 and B2 samples are underdoped in which A1 is YBa$_2$Cu$_3$O$_{7-\delta}$ with $T_c$ = 85 K where the $\delta$ is unknown but not $>$ 0.39; this assumption is estimated by comparing the $\rho(T,\delta)$ curves of A1 and B2 as well as the $T_c$ of each sample in which B2 is YBa$_2$Cu$_3$O$_{6.61}$ with $T_c$ = 58.5 K. These fittings indicate a 2D conduction for both samples due to $-\Delta_{PG}$ values obtained from Eq.~\ref{eq:9} (3D), which are meaningless physically as well as the expected non-$T$-linear metallic behavior above $T^*$ from the Eq.~\ref{eq:9} itself.}
\label{fig:1}
\end{figure}

\begin{figure}

\caption {Experimental $\rho(T,x)$ plots for C3 ($\bullet$) and D4 ($\circ$)  samples that have been fitted using Eq.~\ref{eq:15} at $T$ $<$ $T^*$ and Eq.~\ref{eq:16} at $T$ $>$ $T^*$. The respective C3 and D4 samples are Bi$_2$Sr$_2$Ca$_{0.95}$Y$_{0.05}$Cu$_2$O$_8$ and Bi$_2$Sr$_2$Ca$_{0.89}$Y$_{0.11}$Cu$_2$O$_8$. These samples also indicate a 2D conduction. The fitting of $\rho(T,x)$ curve of sample C3 with Eq.~\ref{eq:9} above $T^*$ as shown with $\Box$ lacks the $T$-linear metallic behavior if observed closely and also give $\Delta_{PG}$ $=$ $-$57 K.}
\label{fig:2}
\end{figure}

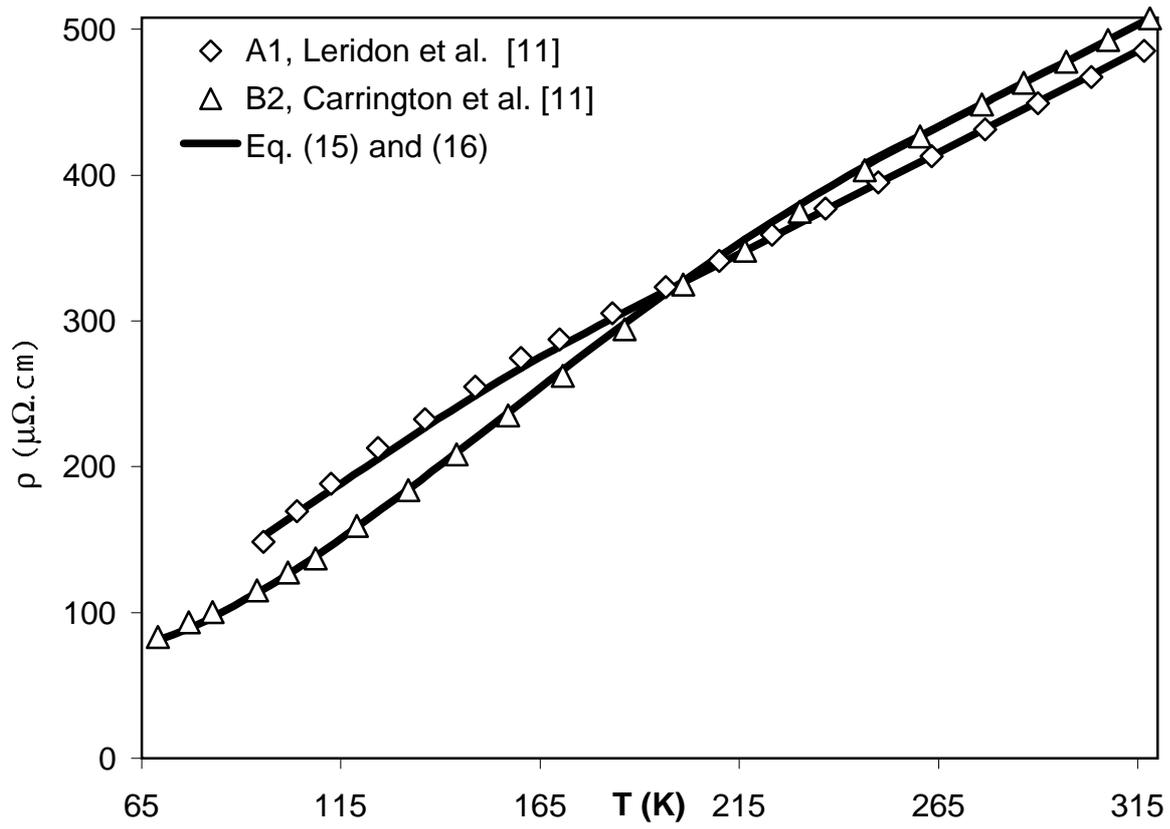
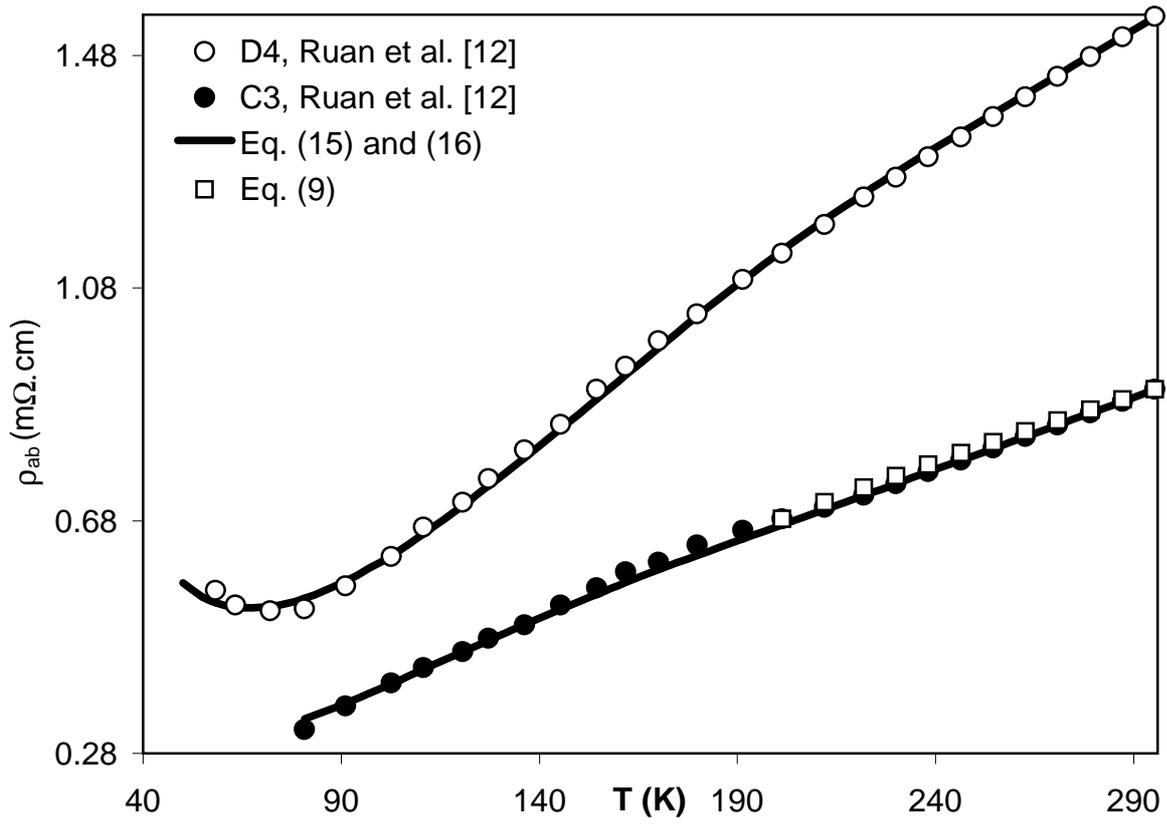
\end{document}